\documentclass[oneside,11pt]{article}
\usepackage{amsmath,amsgen,amstext,amsbsy,amsopn,amsthm, amssymb}
\newtheorem{thm}{Theorem}[section]
\newtheorem{cor}{Corollary}[section]
\newtheorem{lem}{Lemma}[section]

\theoremstyle{definition}
\newtheorem{defn}{Definition}[section]
\newtheorem{rem}{Remark}[section]

\newcommand{\Eqm}{E^{\rm QM}}
\newcommand{\Egp}{E^{\rm GP}}
\newcommand{\Etf}{E^{\rm TF}}
\newcommand{\rtf}{\rho^{\rm TF}}
\newcommand{\rgp}{\rho^{\rm GP}}
\newcommand{\pgp}{\phi^{\rm GP}}
\newcommand{\mtf}{\mu^{\rm TF}}
\newcommand{\mgp}{\mu^{\rm GP}}

\newcommand{\brtf}{\bar\rho}

\newcommand{\x}{{\bf x}}
\newcommand{\0}{{\bf 0}}
\newcommand{\y}{{\bf y}}
\newcommand{\nab}{{\bf\nabla}}

\newcommand{\eps}{\epsilon}
\newcommand{\suli}{\sum\limits}

\newcommand{\V}{V}

\newcommand{\half}{\mbox{$\frac{1}{2}$}}
\newcommand{\E}{\mathcal{E}}

\newcommand{\R}{\mathbb{R}}

\numberwithin{equation}{section}
\begin{document}
\markboth{\scriptsize{LSY 19/10/00}}{\scriptsize{LSY 19/10/00}}
\title{\bf{A Rigorous Derivation of the Gross-Pitaevskii Energy
Functional\\ for
a Two-dimensional Bose Gas}}
\author{\vspace{5pt} Elliott H.~Lieb$^{1}$, Robert Seiringer$^{2}$,
and Jakob
Yngvason$^{2}$\\ \vspace{-4pt}\small{$1.$ Departments of Physics
and Mathematics, Jadwin Hall,} \\ \small{Princeton University,
P.~O.~Box 708, Princeton, New Jersey 08544}\\
\vspace{-4pt}\small{$2.$ Institut f\"ur Theoretische Physik,
Universit\"at Wien}\\ \small{Boltzmanngasse 5, A 1090 Vienna,
Austria}}
\date{\small{ October 19, 2000}  \\
$\phantom{x}$  \\
 Dedicated to Joel Lebowitz on the occasion of his 70th birthday }

\maketitle

\begin{abstract}

We consider the ground state properties of an inhomogeneous
two-dimensional Bose gas with a repulsive, short range pair
interaction and an external confining potential. In the limit when
the particle number $N$ is large but $\bar\rho a^2$ is small,
where $\bar\rho$ is the average particle density and $a$ the
scattering length, the ground state energy and density are
rigorously shown to be given to leading order by a
Gross-Pitaevskii (GP) energy functional with a coupling constant
$g\sim 1/|\ln(\bar\rho a^2)|$. In contrast to the 3D case the
coupling constant depends on $N$ through the mean density.  The GP
energy per particle depends only on $Ng$. In 2D this parameter is
typically so large that the gradient term in the GP energy
functional is negligible and the simpler description by a
Thomas-Fermi type functional is adequate.
\end{abstract}

\renewcommand{\thefootnote}{}
\footnotetext{\copyright 2000 by the authors. Reproduction
of this work, in its entirety, by any means, is permitted for
non-commercial purposes.}

\section{Introduction}

Motivated by recent experimental realizations of Bose-Einstein
condensation the theory of dilute, inhomogeneous Bose gases is
currently a subject of intensive studies.  Most of this work is
based on the assumption that the ground state properties are well
described by the Gross-Pitaevskii (GP) energy functional (see the
review article \cite{DGPS}).  A rigorous derivation of this
functional from the basic many-body Hamiltonian in an appropriate
limit is not a simple matter, however, and has only been achieved
recently for bosons with a short range, repulsive interaction in
three spatial dimensions \cite{LSY2000}.

The present paper is concerned with the justification of the GP
functional in two spatial dimensions.  Several new issues arise.  One
is the form of the nonlinear interaction term in the energy functional
for the GP wave function $\Phi$.  In three dimensions this term is
$4\pi a\int |\Phi|^4$, where $a$ is the scattering length of the
interaction potential.  The rationale is the well known formula for
the energy density of a homogeneous Bose gas, which, for dilute gases
with particle density $\rho$, is $4\pi a\rho^2$.  This fact has
been `known' since the early 50's but a rigorous proof is fairly recent
\cite{LY1998}.
In two dimensions the corresponding formula is
$4\pi\rho^2|\ln(\rho a^2)|^{-1}$ as proved in \cite{LY2000} by
extension of the method of \cite{LY1998}.
The formula was first stated by Schick \cite{schick}; other early
references to this
formula are \cite{hines,popov,fishho,KoSt1992,Ovch}. It would
seem
natural to consider $4\pi\int
|\Phi|^4|\ln(|\Phi|^2 a^2)|^{-1}$ as the interaction term in the GP
functional, and this has indeed been suggested in \cite{Shev,KoSt2000}.
Such a term, however, is unnecessarily complicated for the
purpose of leading order calculations. In fact, since the logarithm
varies only slowly it turns out that
one
can use the {\it same} form as in the three
dimensional case, but with an appropriate dimensionless coupling constant
$g$
replacing the scattering length, and still retain
an exact theory (to leading order in $\rho$).

It is often assumed that a justification of the GP functional depends
on the existence of Bose Einstein condensation. Several remarks can
be made about this: 1. We neither assume nor prove the existence of
BE condensation, but we do demonstrate a kind of condensation over a
distance that is fixed (i.e., non-thermodynamic) but whose length goes
to infinity as the density goes to zero; 2. BE condensation does not
exist in two dimensions when the temperature is positive, but it can,
and most likely does, exist in the ground state; 3. In any event, when
the density is low and the temperature is zero it appears to be likely
that
the system can be described for many purposes in terms of only a few
macroscopic order parameters such as the density and phase -- at least
this is true for the dependence of the ground state energy and density
upon an external potential.

The functional we shall consider is
\begin{equation}\label{gpfunct}
\E^{\rm GP}[\Phi]=\int\left(|\nab\Phi(\x)|^2+\V(\x)|\Phi(\x)|^2+4\pi
g|\Phi(\x)|^4\right){\rm d^2}\x,
\end{equation}
where $V$ is the external
confining potential and all integrals are over $\R^2$.

The choice of $g$ is an issue on which there has not been unanimous
opinion in the recent papers \cite{KoSt2000,KimWon,KimWon2,Garcia,Gonzalez,
Heinrichs,Bayindir}
on this subject.  We shall prove that
a right choice is $g=|\ln (\bar\rho a^2)|^{-1}$ where $\bar\rho$ is a
mean density that will be defined more precisely below.  This mean
density depends on the particle number $N$, which implies that the
scaling properties of the GP functional are quite different in two and
three dimensions.  In the three-dimensional case the natural parameter
is $Na/a_{\rm osc}$, with $a_{\rm osc}$ being the length scale defined by
the external confining potential.  If $a/a_{\rm osc}$ is scaled like
$1/N$ as $N\to\infty$ this parameter is fixed and the gradient term
$\int|\nab
\Phi|^2$ in the GP functional is of the same order as the other terms.
In two dimensions the corresponding parameter is $N|\ln (\bar\rho
a^2)|^{-1}$.  For a quadratic external potential $\bar\rho$ behaves
like $N^{1/2}/a_{\rm osc}^2$ and hence the parameter can only be kept
fixed if $a/a_{\rm osc}$ decreases exponentially with $N$.  A slower
decrease means that the parameter tends to infinity. This corresponds
to the so-called Thomas Fermi (TF) limit where the gradient term has been
dropped altogether and the functional is
\begin{equation}\label{tffunct}
\E^{\rm TF}[\rho]=\int\left(\V(\x)\rho(\x)+4\pi g\rho(\x)^2\right){\rm d^2}\x,
\end{equation}
defined for nonnegative functions $\rho$. Our main result, stated
in Theorems \ref{thm13} and \ref{thm14} below,  is that
minimization of (\ref{tffunct}) reproduces correctly the ground
state energy and density of the many-body Hamiltonian in the limit
when $N\to\infty$, $\bar \rho a^2\to 0$, but $N|\ln (\bar\rho
a^2)|^{-1}\to \infty$. Only in the exceptional situation that
$N|\ln (\bar\rho a^2)|^{-1}$ stays bounded is there need for the
full GP functional (\ref{gpfunct}), cf. Theorems \ref{thm11} and
\ref{thm12}.

We shall now describe the setting more precisely.
The starting point is the Hamiltonian for $N$ identical bosons in an
external
potential $V$ and with pair interaction $v$,
\begin{equation}\label{ham}
H^{(N)}=\suli_{i=1}^{N}\left(-\nab_i^2+\V(\x_i)\right)
+\suli_{i<j}v(\x_i-\x_j),
\end{equation}
acting on the totally symmetric wave functions in
$\otimes^N L^2(\R^{2})$. Units have been chosen so that $\hbar=2m=1$, where $m$ is
the particle mass. We assume that $v$ is nonnegative and
spherically symmetric with a
finite scattering length $a$. (For the definition of scattering length
in two dimensions see the appendix.) The external potential should
be continuous and tend to $\infty$ as
$|\x|\to\infty$. It is then possible and
convenient to shift the energy scale so that
$\min_{\x}V(\x)=0$. For the TF limit theorem we shall require some
additional properties of $V$ to be specified later.

The ground state energy $\omega$ of the one-particle operator $-\nab^2+V$
is a natural energy unit and gives rise to the length unit $a_{\rm
osc}\equiv\omega^{-1/2}$.  In the sequel we shall be considering a limit
where $a/a_{\rm osc}$ tends to zero while $N\to\infty$.  Experimentally
$a/a_{\rm osc}$ can be changed in two ways:  One can either vary $a_{\rm
osc}$ or $a$.  The first alternative is usually simpler in practice but
very recently a direct tuning of the scattering length itself has also
been shown to be feasible \cite{Cornish}.  Mathematically, both
alternatives
are equivalent, of course.  The first corresponds to writing
$V(\x)=a_{\rm osc}^{-2} \hat V(\x/a_{\rm osc})$ and keeping $\hat V$ and
$v$ fixed.  The second corresponds to writing the interaction potential
as $v(\x)=a^{-2}\hat v(\x/a)$, where $\hat v$ has unit scattering length, 
and keeping $V$ and $\hat v$ fixed. This is equivalent to the first, 
since for given $\hat V$ and $\hat v$ the ground state energy of (\ref{ham}),
measured 
in units of $\omega$, depends only on $N$ and $a/a_{\rm osc}$. 
In the dilute limit when $a$ is much smaller than the mean particle 
distance, the energy becomes independent of $\hat v$.

We shall measure all energies in terms of $\omega$ and lengths in 
terms of $a_{\rm osc}$ and regard $\hat V$ and $\hat v$ as fixed. The 
notation $E^{\rm QM}(N,a)$ for the ground state energy of (\ref{ham}) is
then 
justified.

The quantum mechanical particle density is defined by
\begin{equation} \rho^{\rm QM}_{N,a}(\x)=N\int
|\Psi^{(N)}(\x,\x_2,\dots,\x_N)|^2{\rm d^2}\x_2\dots {\rm d^2}\x_N,
\end{equation}
where $\Psi^{(N)}$ is a ground state for (\ref{ham}).

The GP functional (\ref{gpfunct}) has an obvious domain of definition (cf.
Eq.\ (2.1) in \cite{LSY2000}). The infimum of $\E^{\rm GP}[\Phi]$ under the
condition
$\int|\Phi|^2=N$ will be denoted by $\Egp(N,g)$.
The infimum is obtained for a unique, positive function, denoted $\Phi^{\rm
GP}_{N,g}$, and the GP density is defined as $\rho^{\rm
GP}_{N,g}(\x)=\Phi^{\rm
GP}_{N,g}(\x)^2$.

The ground state energy of the TF functional (\ref{tffunct}) with the
subsidiary
condition $\int\rho=N$ is denoted $\Etf(N,g)$. The
corresponding minimizer can be written explicitly; it is
\begin{equation}\label{tfminim}
\rho^{\rm TF}_{N,g}(\x)=\frac 1{8\pi g}[\mu^{\rm TF}-V(\x)]_+,
\end{equation}
where  $[t]_+\equiv\max\{t,0\}$ and $\mu^{\rm TF}$ is chosen so that the
normalization condition $\int \rho^{\rm TF}_{N,g}=N$ holds.

We now define the mean density $\bar\rho$ as the average of
the TF density $\rho^{\rm TF}_{N,1}$ at coupling
constant $g=1$, weighted with $N^{-1}
\rho^{\rm TF}_{N,1}$, i.e.,
\begin{equation}\label{meandens}
    \bar\rho=\frac1N\int\rho^{\rm TF}_{N,1}(\x)^2 {\rm d^2}\x.
\end{equation}
It is clear that $\bar\rho$ depends on $N$ and when we wish to
emphasize this we write $\bar\rho_{N}$.
The definition (\ref{meandens}) has the advantage that $\bar\rho$ is easily
computed; for instance, if $V(\x)\sim |\x|^s$ for some $s>0$,
then $\bar\rho_{N}\sim N^{s/(s+2)}$. It may appear more
natural to define $\bar\rho$ self-consistently as
$\bar\rho=\frac1N\int\rho^{\rm
TF}_{N,g}(\x)^2 {\rm d^2}\x$ with $g=|\ln (\bar\rho a^2)|^{-1}$, which amounts
to solving a nonlinear equation for $\bar\rho$. Also, the TF density
could  be replaced by the GP density.
However, since $\bar\rho$ will only appear under a logarithm
such sophisticated definitions are not needed for
the leading order result we are after. The simple formula
(\ref{meandens}) is adequate for our purpose, but it should be kept in
mind that the self-consistent
definition may be relevant in computations beyond the leading order.

With this notation we can now state the two dimensional analogue
of Theorem I.1 in \cite{LSY2000}.
\begin{thm}[GP limit for the energy]\label{thm11}
If, for $N\to\infty$, $a^2\brtf_N\to 0$ with
$N/|\ln(a^2\brtf_N)|$ fixed, then
\begin{equation}
\lim_{N\to\infty}\frac{\Eqm(N,a)}{\Egp(N,1/|\ln(a^2\brtf_N)|)}=
1.
\end{equation}
\end{thm}
The corresponding theorem for the density, c.f.\ Theorem I.2 in
\cite{LSY2000},
is
\begin{thm}[GP limit for the density]\label{thm12}
If, for $N\to\infty$, $a^2\brtf_N\to 0$ with
$\gamma\equiv N/|\ln(a^2\brtf_N)|$ fixed, then
\begin{equation}
\lim_{N\to\infty}\frac1N\rho^{\rm QM}_{N,a}(\x)=\rho^{\rm GP}_{1,\gamma}(\x)
\end{equation}
in the sense of weak convergence in $L^1(\R^2)$.
\end{thm}
These theorems, however, are not particularly useful in the two dimensional
case, because the hypothesis that $N/|\ln(a^2\brtf_N)|$ stays bounded
requires
an exponential decrease of $a$ with $N$.
As remarked above, the TF limit, where $N/|\ln(a^2\brtf_N)|\to\infty$, is
much
more relevant. Our treatment of this limit requires that $V$ is
asymptotically
homogeneous and
sufficiently regular in a sense made precise below. This condition can be
relaxed, but it seems adequate for most
practical applications and simplifies things considerably.
\begin{defn} We say that $V$ is {\it asymptotically homogeneous} of
order $s>0$ if there is a function $W$ with $W(\x)\neq 0$ for
$\x\neq \0$ such that
\begin{equation}\label{asymp}
\frac{\lambda^{-s}V(\lambda \x)-W(\x)}{1+|W(\x)|}\to 0\quad {\rm as} \quad
\lambda\to\infty
\end{equation}
and the convergence is uniform in $\x$.
\end{defn}
The function $W$ is clearly uniquely determined and homogeneous of
order $s$, i.e., $W(\lambda \x)=\lambda^sW(\x)$ for all $\lambda\geq 0$.
\begin{thm}[TF limit for the energy]\label{thm13}
Suppose $V$ is asymptotically homogeneous of order $s>0$ and its scaling
limit
$W$ is locally
H\"older continuous, i.e., $|W(\x)-W(\y)|\leq {\rm (const.)}|\x-\y|^\alpha$
for
$|\x|,|\y|=1$ for some fixed $\alpha>0$. If, for $N\to\infty$,
$a^2\brtf_N\to
0$
but $N/|\ln(a^2\brtf_N)|\to\infty$, then
\begin{equation}
\lim_{N\to\infty}\frac{\Eqm(N,a)}{\Etf(N,1/|\ln(a^2\brtf_N)|)}=
1.
\end{equation}
\end{thm}
To state the corresponding theorem for the density we need the
minimizer of (\ref{tffunct}) with $g=1$, $V$ replaced by $W$, and
normalization $\int\rho=1$. We shall denote this minimizer by
$\tilde\rho^{\rm TF}_{1,1}$; an explicit formula is
\begin{equation}\label{rhotilde}
\tilde\rho^{\rm TF}_{1,1}(\x)=\frac 1{8\pi}[\tilde\mu^{\rm
TF}-W(\x)]_+,
\end{equation}
where $\tilde\mu^{\rm TF}$ is determined by the normalization
condition.
\begin{thm}[TF limit for the density]\label{thm14}
Let $V$ satisfy the same hypothesis as in Theorem \ref{thm13}. If,
for $N\to\infty$, $a^2\brtf_N\to 0$ but
$\gamma=N/|\ln(a^2\brtf_N)|\to\infty$, then
\begin{equation}
\lim_{N\to\infty}\frac{\gamma^{2/(s+2)}}N\rho^{\rm
QM}_{N,a}(\gamma^{1/(s+2)}\x) =\tilde\rho^{\rm TF}_{1,1}(\x)
\end{equation}
in the sense of weak convergence in $L^1(\R^2)$.
\end{thm}
\begin{rem}
For large $N$, $\bar\rho_N$ behaves like ${\rm
(const.)}N^{s/(s+2)}$. Moreover, prefactors are unimportant in the
limit $N\to \infty$, because $\bar\rho_N$ stands under a
logarithm. Hence Theorems \ref{thm13} and \ref{thm14} could also
be stated with $N^{s/(s+2)}$ in place of $\bar\rho_N$.
\end{rem}
The proofs of these theorems follow from upper and lower bounds on the
ground
state energy $E^{\rm QM}(N,a)$ that are derived in Sections \ref{sect3} and
\ref{sect4}. For
these
bounds some properties of the minimizers of the functionals (\ref{gpfunct})
and
(\ref{tffunct}), discussed in the following section, are needed.

\section{GP and TF theory}

In this section we consider the functionals (\ref{gpfunct}) and
(\ref{tffunct})
with an arbitrary posi\-tive coupling constant $g$. Existence and
uniqueness of
minimizers is shown in the same way as in Theorem II.1 in \cite{LSY2000}.
The GP energy $\Egp(N,g)$  has the simple
scaling property $\Egp(N,g)=N\Egp(1,Ng)$. Likewise, $N^{-1/2} \Phi^{\rm
GP}_{N,g}\equiv \phi^{\rm GP}_{\gamma}$ depends only
on
\begin{equation}\gamma\equiv Ng
    \end{equation}and satisfies the normalization condition $\int
|\phi^{\rm
GP}_{\gamma}|^2=1$.  The variational equation (GP equation) for the GP
minimization problem, written in terms of
$\phi^{\rm GP}_{\gamma}$, is
\begin{equation}\label{gpeq}
-\Delta\pgp_\gamma+V\pgp_\gamma+8\pi
\gamma(\pgp_\gamma)^3=\mgp(\gamma)\pgp_\gamma,
\end{equation}
where the Lagrange multiplier (chemical
potential) $\mgp(\gamma)$ is determined by the subsidiary normalization
condition.
Multiplying (\ref{gpeq}) with $\phi^{\rm GP}_{\gamma}$ and integrating we
obtain
\begin{equation}\label{mugp}
\mu^{\rm GP}(\gamma)=\Egp(1,\gamma)+4\pi \gamma\int\phi^{\rm
GP}_{\gamma}(\x)^4{\rm d}^2\x.
\end{equation}

For the upper bound on the quantum mechanical energy in the next section we
shall need a bound on the absolute value of the minimizer $\pgp_\gamma$.

\begin{lem}[Upper bound for the GP minimizer]\label{gpbound}
\begin{equation}
\|\pgp_\gamma\|_\infty^2\leq\frac{\mgp(\gamma)}{8\pi \gamma}
\end{equation}
\end{lem}
\begin{proof}
$\pgp_\gamma$ is a continuous and positive function that satisfies the
variational
equation
\begin{equation}
-\Delta\pgp_\gamma+U\pgp_\gamma=\mgp\pgp_\gamma
\end{equation}
with $U=V+8\pi \gamma(\pgp_\gamma)^2$. Let ${\cal B}=\{\x\,| \,
\pgp_\gamma(\x)^2>\mgp/(8\pi\gamma)\}$. Since $V\geq 0$ we see that
$-\Delta\pgp_\gamma\leq 0$ on ${\cal B}$, i.e., $\pgp_\gamma$ is subharmonic on  
${\cal B}$. Hence $\pgp_\gamma$ achieves its
maximum on the boundary of ${\cal B}$, where
$\pgp_\gamma(\x)^2=\mgp/(8\pi\gamma)$, so ${\cal B}$ is empty.
\end{proof}

The ground state energy $\Etf(N,g)$ of the TF functional (\ref{tffunct})
scales
in the same way as $\Egp(N,g)$, i.e.,
$\Etf(N,g)=N\Etf(1,Ng)$, and the
corresponding minimizer $\rho^{\rm TF}_{N,g}$ is equal to $N\rho^{\rm
TF}_{1,Ng}$. For short, we shall denote $\rho^{\rm TF}_{1,\gamma}$ by
$\rho^{\rm
TF}_\gamma$. By (\ref{tfminim}) we have
\begin{equation}
\rho^{\rm TF}_\gamma(\x)=\frac 1{8\pi \gamma}[\mu^{\rm
TF}(\gamma)-V(\x)]_+,
\end{equation}
with the chemical potential $\mu^{\rm TF}(\gamma)$ determined by the
normalization condition $\int\rtf_\gamma=1$. In the same way as in
(\ref{mugp})
we have
\begin{equation}\label{mutf}
\mu^{\rm TF}(\gamma)=\Etf(1,\gamma)+4\pi \gamma\int\rho^{\rm
TF}_{\gamma}(\x)^2{\rm d}^2\x.
\end{equation}
The chemical potential can also be computed from a variational principle:

\begin{lem}[Variational principle for $\mtf$]\label{lemvarmu}
\begin{equation}\label{varmu}
\mtf(\gamma)=\inf_{\rho\geq 0, \int\rho=1}\int V\rho+8\pi
\gamma\|\rho\|_\infty
\end{equation}
\end{lem}
\begin{proof}
Obviously, the infimum is achieved for a multiple of a characteristic
function
for some measurable set ${\cal R}\subset\R^2$. If $|{\cal R}|$ denotes
the 
Lebesgue measure of $\cal R$, then
\begin{eqnarray}\nonumber
& &\inf_{\int\rho=1}\int V\rho+8\pi \gamma\|\rho\|_\infty\\& &=\inf_{\cal
R}\left(
\int_{\cal R} V+8\pi \gamma\right)\frac 1{|{\cal R}|}\\
& &=\inf_{\cal R}\left(\int_{\cal R}\left(V-\mtf(\gamma)\right)+8\pi
\gamma+\mtf(\gamma)
{|{\cal R}|}\right)\frac 1{{|{\cal R}|}}.
\end{eqnarray}
Now $\int_{\cal R}(V-\mtf(\gamma))\geq -8\pi \gamma$, with equality for
\begin{equation}
\left\{\x|V(\x)<\mtf(\gamma)\right\}\subseteq {\cal R}\subseteq
\left\{\x|V(\x)\leq\mtf(\gamma)\right\}.
\end{equation}
\end{proof}

\begin{cor}[Properties of $\mtf(\gamma)$]\label{propmu}
$\mtf(\gamma)$ is a concave and mono\-tonously increasing function
of $\gamma$ with $\mtf(0)=0$. Hence $\mtf(\gamma)/\gamma$ is
decreasing in $\gamma$. Moreover, $\mtf(\gamma)\to\infty$ and
$\mtf(\gamma)/\gamma\to 0$ as $\gamma\to\infty$.
\end{cor}
\begin{proof}
Immediate consequences of Lemma \ref{lemvarmu}, using that
$\min_\x V(\x)=0$ and $\lim_{|\x|\to\infty}V(\x)=\infty$.
\end{proof}

Note that since $\Etf(1,\gamma)\geq \half\mtf(\gamma)$ we also see that
$\Etf(1,\gamma)\to\infty$ with $\gamma$.
In this limit the GP energy converges to the
TF energy, provided the external potential satisfies a mild
regularity and growth condition:

\begin{lem}[TF limit of the GP energy]
Suppose for some constants $\alpha>0$, $L_1$ and $L_2$
\begin{equation}\label{cond1}
|V(\x)-V(\y)|\leq L_1|\x-\y|^\alpha e^{L_2|\x-\y|}(1+V(\x)).
\end{equation}
Then
\begin{equation}\label{gptotf}
\lim_{\gamma\to\infty}\frac{\Egp(1,\gamma)}{\Etf(1,\gamma)}=1.
\end{equation}
\end{lem}
\begin{proof}
It is clear that $\Etf(1,\gamma)\leq \Egp(1,\gamma)$. For the other
direction, we use $(j_\eps*\rtf_\gamma)^{1/2}$ as a test function for
$\E^{\rm GP}$, where
\begin{equation}
j_\eps(\x)=\frac 1{2\pi\eps^2}\exp\left(-\frac 1\eps |\x|\right).
\end{equation}
Note that $\int j_\eps=1$ and $|\nab j_\eps|=\eps^{-1}j_\eps$.
Therefore
\begin{eqnarray}\nonumber
\Egp(1,\gamma)&\leq&\int\left(\frac{1}{4 j_\eps*\rtf_\gamma}|\nab
j_\eps*\rtf_\gamma|^2+V(j_\eps*\rtf_\gamma)+4\pi
\gamma(j_\eps*\rtf_\gamma)^2\right)\\
&\leq&\frac1{4\eps^2}+\int\left((j_\eps*V)\rtf_\gamma+4\pi
\gamma(\rtf_\gamma)^2\right),
\end{eqnarray}
where we have used convexity for the last term. Moreover,
\begin{eqnarray}\nonumber
\int(j_\eps*V-V)\rtf_\gamma &=&\int\int {\rm d^2}\x {\rm d^2}\y
j_\eps(\x-\y)\left(V(\x)-V(\y)\right)\rtf_\gamma(\x)\\ \nonumber
&\leq&\frac{L_1}{2\pi\eps^2} \int\int {\rm d^2}\x {\rm
d^2}\y|\x-\y|^\alpha
e^{(-\eps^{-1}+L_2)|\x-\y|}(1+V(\x))\rtf_\gamma(\x)\\ &\leq&{\rm
(const.)}\, \eps^\alpha\left(1+\Etf(1,\gamma)\right),
\end{eqnarray}
as long as $\eps< L_2^{-1}$. So we have
\begin{equation}
\Egp(1,\gamma)\leq (1+{\rm (const.)}\, \eps^\alpha)\Etf(1,\gamma)+\frac
1{4\eps^2}+{\rm (const.)}\, \eps^\alpha.
\end{equation}
Optimizing over $\eps$ gives as a final result
\begin{equation}
\Egp(1,\gamma)\leq \Etf(1,\gamma)\left(1+{\rm
(const.)}\Etf(1,\gamma)^{-\alpha/(\alpha+2)}\right).
\end{equation}
\end{proof}

Condition (\ref{cond1}) is in particular fulfilled if $V$ is
homogeneous of some order $s>0$ and locally H\"older continuous.
In this case,
\begin{equation}
\Etf(1,\gamma)=\gamma^{s/(s+2)}\Etf(1,1)
\end{equation}
and
\begin{equation}
\gamma^{2/(s+2)}\rtf_\gamma(\gamma^{1/(s+2)}\x)=\rho^{\rm TF}_{1,1}(\x).
\end{equation}
By (\ref{mutf}) we also have
\begin{equation}
\mtf(\gamma)=\gamma^{s/(s+2)}\mtf(1).
\end{equation}

If $V$ is asymptotically homogeneous
with a locally
H\"older continuous limiting
function $W$, we can prove corresponding formulas for the limit
$\gamma\to\infty$. This is the content of the next theorem, where we have
included results on the GP $\to$ TF limit as well:
\begin{thm}[Scaling limits]\label{tildeetf}
Suppose $V$ satisfies the condition of Theorem \ref{thm13}. Let
$\tilde\Etf(1,1)$ be the minimum of the TF functional
(\ref{tffunct}) with $g=1$ and $N=1$ and $V$ replaced by $W$, and
let $\tilde\rtf_{1,1}$ be the corresponding minimizer. Then
\begin{enumerate}
\item[{\rm (i)}]$\lim_{\gamma\to\infty}\Egp(1,\gamma)/\gamma^{s/(s+2)}=
\lim_{\gamma\to\infty}\Etf(1,\gamma)/\gamma^{s/(s+2)}=\tilde\Etf(1,1)$.
\item[{\rm(ii)}]$\lim_{\gamma\to\infty}\gamma^{2/(s+2)}\rgp_{1,\gamma}
(\gamma^{1
/(s+2)}\x)=\tilde\rtf_{1,1}(\x)$, strongly in $L^2(\R^2)$.
\item[{\rm(iii)}]$\lim_{\gamma\to\infty}\gamma^{2/(s+2)}\rtf_{\gamma}
(\gamma^{
1/(s+2)}\x)=\tilde\rtf_{1,1}(\x)$, uniformly in $\x$.
\end{enumerate}
\end{thm}
\begin{proof}
With the demanded properties of $V$, (\ref{gptotf}) holds. Using
this and (\ref{asymp}) one easily verifies (i). Moreover,
$\gamma^{2/(s+2)}\rgp_{1,\gamma}(\gamma^{1 /(s+2)}\x)$ is a
minimizing sequence for the functional in question, so we can
conclude as in Theorem II.2 in \cite{LSY2000} that it converges to
$\tilde\rtf_{1,1}(\x)$ strongly in $L^2$, proving (ii). (Remark: In Eq.\
(2.10)
in \cite{LSY2000} there is a misprint, instead of $\rgp_{1,Na}$ one should
have
$\tilde \rgp_{1,Na}$ on the left side.) To see
(iii) let us define
\begin{equation}\label{rhohat}
\widehat\rho_\gamma(\x)=\gamma^{2/(s+2)}
\rtf_\gamma\left(\gamma^{1/(s+2)}\x\right).
\end{equation}
We can write
\begin{equation}\label{rhat}
\widehat\rho_\gamma(\x)=\frac 1{8\pi}
\left[\gamma^{-s/(s+2)}\mtf(\gamma)-W(\x)-\eps(\gamma,\x)\right]_+
\end{equation}
with
\begin{equation}
\eps(\gamma,\x)=\gamma^{-s/(s+2)}V(\gamma^{1/(s+2)}\x)- W(\x).
\end{equation}
By assumption, $|\eps(\gamma,\x)|<\delta(\gamma)(1+W(\x))$ for some
$\delta(\gamma)$ with $\lim_{\gamma\to\infty}\delta(\gamma)=0$.
Because $\int\widehat\rho_\gamma=1$ for all $\gamma$, we see from Eq.\
(\ref{rhat}) that
$\mtf(\gamma)\gamma^{-s/(s+2)}$ converges to some $c$ as
$\gamma\to\infty$. Moreover, we can conclude that the support of
$\widehat\rho_\gamma$ is for large $\gamma$ contained in some bounded set
${\cal
B}$
independent of $\gamma$. Therefore
\begin{equation}
1=\lim_{\gamma\to\infty}\int\widehat\rho_\gamma=\int
(8\pi)^{-1}[c-W(\x)]_+
\end{equation}
by dominated convergence, so $c$ is equal to the $\tilde\mtf$ of Eq.\
(\ref{rhotilde}). Now
\begin{equation}\label{hatrho}
\widehat\rho_\gamma(\x)=\frac 1{8\pi}
\left[\tilde\mtf-W(\x)-\bar\eps(\gamma,\x)\right]_+
\end{equation}
with
\begin{equation}
\bar\eps(\gamma,\x)=\eps(\gamma,\x)+\tilde\mtf-\gamma^{-s/(s+2)}\mtf(\gamma
).
\end{equation}
Again $|\bar\eps(\gamma,\x)|<\bar\delta(\gamma)(1+W(\x))$ for some
$\bar\delta(\gamma)$ with
$\lim_{\gamma\to\infty}\bar\delta(\gamma)=0$. By Eqs.\ (\ref{rhotilde}) and
(\ref{hatrho}) we thus have
\begin{equation}
\|\widehat\rho_\gamma-\tilde\rtf_{1,1}\|_\infty<C\bar\delta(\gamma).
\end{equation}
with $C=(8\pi)^{-1}\sup_{\x\in{\cal B}}(1+W(\x))<\infty$.
\end{proof}

The {\it mean density} for the TF theory is defined by
\begin{equation}
\brtf_\gamma\equiv N\int\rtf_\gamma(\x)^2{\rm d}^2\x.
\end{equation}
For $\gamma=N$, i.e., $g=1$ this is the same as
(\ref{meandens}). It satisfies

\begin{lem}[Bounds on $\brtf_\gamma$]\label{rhobarbound}
For some constant $C>0$
\begin{equation}
N\frac{\mtf(\gamma)}{8\pi\gamma}\geq\brtf_\gamma\geq
CN\frac{\mtf(\gamma)}\gamma.
\end{equation}
\end{lem}
\begin{proof}
The upper bound is trivial. Because $\widehat\rho_\gamma$, defined
in (\ref{rhohat}), converges uniformly to $\tilde\rtf_{1,1}$ and
$\mtf(\gamma)\gamma^{-s/(s+2)}\to\tilde\mtf$ as $\gamma\to\infty$,
we have the lower bound
\begin{equation}
\frac{\gamma\brtf_\gamma}{N\mtf(\gamma)}\geq
8\pi\gamma^{s/(s+2)}\mtf(\gamma)^{-1}\left(\int(\tilde\rtf_{1,1})^2
-2\|\tilde\rtf_{1,1}-\widehat\rho\|_\infty\right)>C
\end{equation}
for some $C>0$.
\end{proof}

\begin{rem}\label{rem21}
With $V$ asymptotically homogeneous of order $s$,
$\mtf(\gamma)\gamma^{-s/(s+2)}$ converges as $\gamma\to\infty$,
i.e. $\mtf(\gamma)\sim \gamma^{s/(s+2)}$ for large $\gamma$. So the
mean TF density for coupling
constant $g=1$, defined in (\ref{meandens}), has the asymptotic behavior
$\brtf\sim N^{s/(s+2)}$.
\end{rem}

\section{Upper bound to the QM energy}\label{sect3}

As in the three dimensional case, cf.\ Eqs.\ (3.29) and (3.27) in
\cite{LSY2000}, one has the upper bound
\begin{equation}
\frac{\Eqm(N,a)}N\leq
\frac{\int|\nab\pgp_\gamma|^2+V(\pgp_\gamma)^2}{1-N\|\pgp_\gamma\|_\infty
^2I}+
\frac{NJ\int(\pgp_\gamma)^4+\mbox{$\frac
23$}N^2(\|\pgp_\gamma\|_\infty^2K)^2} {(1-N\|\pgp_\gamma\|_\infty^2I)^2},
\end{equation}
where we have implicitly used that $-\Delta\pgp_\gamma+V\pgp_\gamma\geq 0$,
which is justified by Lemma \ref{gpbound}. The coefficients $I$, $J$ and
$K$ are
given by Eqs.\ (2.4)--(2.10) in \cite{LY2000}. They depend on the
scattering
length
and a parameter $b$. We choose
$\gamma=N/|\ln(a^2\brtf)|$ and $b=\brtf^{-1/2}$. (Recall that $\bar\rho$ is
short for $\bar\rho_N$.) With this choice
we have (as long as
$a^2\brtf<1$)
\begin{equation}
J=\frac{4\pi}{|\ln(a^2\brtf)|},
\end{equation}
and the error terms
\begin{equation}
N\|\pgp_\gamma\|_\infty^2 I\leq {\rm (const.)}\frac{\mgp(\gamma)}{\brtf}
\left(1+
O(|\ln(a^2\brtf)|^{-1})\right)
\end{equation}
and
\begin{equation}
K^2N^2\|\pgp_\gamma\|_\infty^4\leq {\rm (const.)}\Egp(1,\gamma)\frac
{\mgp(\gamma)}{\brtf}
\left(1+O(|\ln(a^2\brtf)|^{-1})\right),
\end{equation}
where we have used Lemma \ref{gpbound}. So we have the upper bound
\begin{equation}
\frac{\Eqm(N,a)}{\Egp(N,1/|\ln(a^2\brtf)|)}\leq
1+O\left(\mgp(\gamma)/\brtf)
+O((|\ln(a^2\brtf)|^{-1})\right).
\end{equation}
Now if $\gamma$ is fixed as $N\to\infty$
\begin{equation}
\frac{\mgp(\gamma)}{\brtf}\sim\frac{1}{|\ln(a^2\brtf)|}\sim
\frac 1N.
\end{equation}
If $\gamma\to\infty$ with $N$ we have instead, assuming that the
external potential is asymptotically homogeneous of order $s$,
\begin{equation}
\frac{\mgp(\gamma)}{\brtf}\sim\frac{\mtf(\gamma)}{\mtf(N)}\sim
\left(\frac \gamma N\right)^{s/(s+2)},
\end{equation}
so in any case
\begin{equation}\label{upper}
\frac{\Eqm(N,a)}{\Egp(N,1/|\ln(a^2\brtf)|)}\leq
1+O\left(|\ln(a^2\brtf)|^{-s/(s+2)}\right)
\end{equation}
holds as $N\to\infty$ and $a^2\brtf\to 0$.

\section{Lower bound to the QM energy}\label{sect4}

Compared to the treatment of the 3D problem in \cite{LSY2000} the new issue
here is
the TF case, i.e., $\gamma=N/|\ln(a^2\brtf)|\to\infty$, and we discuss this
case
first.  The GP limit with $\gamma$ fixed can be treated in complete analogy
with
the 3D case, cf.\ Remark \ref{rem41} below.

We introduce again the
rescaled $\widehat\rho_\gamma$ as in \eqref{rhohat} and also
\begin{equation}
\widehat v(\x)=\gamma^{2/(s+2)}\, v\left(\gamma^{1/(s+2)}\x\right).
\end{equation}
Note that the scattering length of $\widehat v$ is $\widehat
a=a\,\gamma^{-1/(s+2)}$. Using $V\geq\mtf(\gamma)-8\pi
\gamma\rtf_\gamma$ and (\ref{mutf}) we see that
\begin{eqnarray}\nonumber
\Eqm(N,a)&\geq&\Etf(N,\gamma/N)+4\pi N\gamma^{s/(s+2)}\int\widehat
\rho_\gamma^2 +\gamma^{-2/(s+2)}Q\\ & &-8\pi
N\gamma^{s/(s+2)}\|\widehat\rho_\gamma-\tilde\rtf_{1,1}\|_\infty,
\end{eqnarray}
with
\begin{equation}
Q=\inf_{\int|\Psi|^2=1}\sum_{i}\int\left(|\nab_i\Psi|^2+\sum_{j<i}
\widehat v(\x_i-\x_j)|\Psi|^2-8\pi
\gamma\tilde\rtf_{1,1}(\x_i)|\Psi|^2\right).
\end{equation}
Dividing space into boxes $\alpha$ of side length $L$ with Neumann
boundary conditions we get
\begin{equation}\label{box}
Q\geq \sum_\alpha E^{\rm hom}(n_\alpha,L)-8\pi
\gamma\rho_{\alpha,\max}n_\alpha,
\end{equation}
where $\rho_{\alpha,\max}$ denotes the maximal value of
$\tilde\rtf_{1,1}$ in the box $\alpha$, and $E^{\rm hom}(n,L)$ is
the energy of a homogeneous gas of $n$ bosons in a box of side
length $L$ and Neumann boundary conditions. We can forget about
the boxes where $\rho_{\alpha,\max}=0$, because the energy of
particles in these boxes is positive.

We now want to use the lower bound on $E^{\rm hom}$ given in
\cite{LY2000}, namely
\begin{equation}\label{ehom}
E^{\rm hom}(n,L)\geq 4\pi \frac{n^2}{L^2}\frac 1{|\ln(\widehat
a^2n/L^2)|} \left(1-C|\ln(\widehat a^2n/L^2)|^{-1/5}\right).
\end{equation}
This bound holds for $n>{\rm (const.)} |\ln(\widehat a^2
n/L^2)|^{1/5}$ and small enough $\widehat a^2 n/L^2$. Now if the
minimum in (\ref{box}) is taken in some box $\alpha$ for some
value $n_\alpha$, we have
\begin{equation}
E^{\rm hom}(n_\alpha+1,L)-E^{\rm hom}(n_\alpha,L)\geq
8\pi\gamma\rho_{\alpha,\max}.
\end{equation}
By a computation analogous to the upper bound (see \cite{LSY2000})
one shows that
\begin{eqnarray}\nonumber
& &E^{\rm hom}(n+1,L)-E^{\rm hom}(n,L)\\ \label{chempot} & &\leq
8\pi\frac n{L^2}\frac 1{|\ln(\widehat a^2
n/L^2)|}\left(1+O\left(|\ln(\widehat a^2
n/L^2)|^{-1}\right)\right).
\end{eqnarray}
Using Lemma \ref{rhobarbound} and the asymptotics of $\mtf$ (Remark
\ref{rem21}) we
see that
\begin{equation}\label{seethat}
\frac{\widehat a^2 n}{L^2}\leq \frac{\widehat a^2
N}{L^2}=N^{s/(s+2)}\left(\frac
N\gamma\right)^{2/(s+2)}\frac{a^2}{L^2}\leq
a^2\brtf\frac{C}{L^2}\left(\frac N\gamma\right)^{2/(s+2)},
\end{equation}
for some constant $C$, so (\ref{chempot}) reads
\begin{eqnarray}\nonumber
& &E^{\rm hom}(n+1,L)-E^{\rm hom}(n,L)\\ \label{chempot2} & &\leq
8\pi\frac n{L^2}\frac
1{|\ln(a^2\brtf)|}\left(1+O\left(\frac{1+|\ln((\gamma/N)^{2/(s+2)}L^2/C)|}{
|\ln(a^2 \brtf)|}\right)\right).
\end{eqnarray}
So if $L$ is fixed, our minimizing $n_\alpha$ is at least $\sim
\rho_{\alpha,\max} L^2 N$. If $N$ is large enough and
$a^2\brtf$ is small enough, we can thus use (\ref{ehom}) in
(\ref{box}) to get
\begin{equation}\label{Q1}
Q\geq \sum_\alpha 4\pi \left(\frac{n_\alpha^2}{L^2} \frac
1{|\ln\left(\frac{\widehat a^2n_\alpha}{L^2}\right)|}\left(1-\frac
C {|\ln\left(\frac{\widehat a^2N}{L^2}\right)|^{1/5}}\right)-2
\frac{N\rho_{\alpha,\max}}{|\ln(a^2\brtf)|}\right).
\end{equation}

\begin{lem}\label{xb}
For $0<x,b<1$ we have
\begin{equation}
\frac{x^2}{|\ln x|}-2\frac b{|\ln b|}x\geq -
\frac{b^2}{|\ln b|}\left(1+\frac 1{(2|\ln b|)^2}\right).
\end{equation}
\end{lem}
\begin{proof}
Since $\ln x\geq-\frac 1{de}x^{-d}$ for
all $d>0$ we have
\begin{equation}
\frac{x^2}{b^2}\frac{|\ln b|}{|\ln x|}
-2\frac xb\geq\frac{|\ln b|}{b^2}ed x^{2+d}-\frac{2x}{b}
\geq c(d)(b^ded|\ln b|)^{-1/(1+d)}
\end{equation}
with
\begin{equation}
c(d)=2^{(2+d)/(1+d)}\left(\frac 1{(2+d)^{(2+d)/(1+d)}}-\frac 1
{(2+d)^{1/(1+d)}}\right)\geq -1-\frac 14d^2.
\end{equation}
Choosing $d=1/|\ln b|$ gives the desired result.
\end{proof}

Note that the Lemma above implies for $k\geq 1$
\begin{equation}
\frac{x^2}{|\ln x|}-2\frac b{|\ln b|}xk\geq -
\frac{b^2}{|\ln b|}\left(1+\frac 1{(2|\ln b|)^2}\right)k^2.
\end{equation}
Applying this with $x=\widehat a^2n_\alpha/L^2$ and $b=N\widehat
a^2 \rho_{\alpha,\max}$ we get the bound
\begin{eqnarray}\nonumber
&Q&\geq-4\pi N\gamma\sum_\alpha \rho_{\alpha,\max}^2L^2\\
\nonumber & &\times\left[\left(1+\frac1{4|\ln(\widehat
a^2N\rho_{\alpha,\max})|^2}\right) \frac{|\ln(\widehat
a^2N\rho_{\alpha,\max})|}{|\ln(a^2\brtf)|} \left(1-\frac C
{|\ln\left(\frac{\widehat
a^2N}{L^2}\right)|^{1/5}}\right)^{-1}\right]\\
\end{eqnarray}
for (\ref{Q1}). To estimate the error terms, note that as in
(\ref{seethat})
\begin{equation}
\widehat a^2N \sim a^2\brtf\left(\frac N\gamma\right)^{2/(s+2)},
\end{equation}
so $|\ln(\widehat a^2 N)|=|\ln(a^2\brtf)|+O(\ln|\ln(a^2\brtf)|)$ for
small $a^2\brtf$. Using
$\|\widehat\rho_\gamma-\tilde\rtf_{1,1}\|_\infty\to 0$ (Theorem
\ref{tildeetf}
(iii)) and
$\int\widehat\rho_\gamma^2\to\int(\tilde\rtf_{1,1})^2$ as
$\gamma\to\infty$ (which follows from the uniform convergence and
boundedness of
the supports) we get
\begin{equation}
\liminf_{N\to\infty}\frac{\Eqm(N,a)}{\Etf(N,1/|\ln(a^2\brtf)|)}\geq
1-{\rm (const.)} \left(\sum_\alpha\rho_{\alpha,\max}^2 L^2
-\int(\tilde\rtf_{1,1})^2\right).
\end{equation}
Since this holds for all choices of the boxes $\alpha$ with
arbitrary small side length $L$, and by the assumptions on $V$
$\tilde\rtf_{1,1}$ is continuous and has compact support, we can
conclude
\begin{equation}\label{lowertf}
\liminf_{N\to\infty}\frac{\Eqm(N,a)}{\Etf(N,1/|\ln(a^2\brtf)|)}\geq
1
\end{equation}
in the limit $N\to\infty$, $a^2\brtf\to 0$ and
$N/|\ln(a^2\brtf)|\to\infty$.

\begin{rem}[The GP case]\label{rem41}
In the derivation of the lower bound we have assumed that
$\gamma\to\infty$ with $N$, i.e. $N\gg |\ln(a^2\brtf)|$, which
seems natural because otherwise the scattering length would have to
decrease exponentially with $N$. However, for fixed $\gamma$ one
can use the methods of \cite{LSY2000} (with slight modifications:
One uses the 2D bounds on the homogeneous gas and Lemma \ref{xb})
to compute a lower bound in terms of the GP energy. The result is
\begin{equation}\label{lowergp}
\liminf_{N\to\infty}\frac{\Eqm(N,a)}{\Egp(N,1/|\ln(a^2\brtf)|)}\geq
1
\end{equation}
in the limit $N\to\infty$, $a^2\brtf\to 0$ with
$\gamma=N/|\ln(a^2\brtf)|$ fixed.
\end{rem}

\section{The limit theorems}

We have now all the estimates needed for Theorems
\ref{thm11}--\ref{thm14}.
The upper bound
(\ref{upper}) and the lower bound (\ref{lowergp}) prove Theorem
\ref{thm11}.
The energy limit Theorem \ref{thm13} for the TF case follows from
(\ref{upper}),
Theorem \ref{tildeetf} (i) and (\ref{lowertf}).

The convergence of the energies implies the convergence of the
densities in the usual way by variation of the external potential.
Replacing
$V(\x)$ by
$V(\x)+\delta\gamma^{s/(s+2)}Y(\gamma^{-1/(s+2)}\x)$ for some
positive $Y\in C_0^\infty$ and redoing the upper and lower
bounds we see that Theorem \ref{thm13} and Theorem
\ref{tildeetf} (i) hold with $W$ replaced by $W+\delta Y$.
Differentiating with respect to $\delta$ at $\delta=0$ yields
\begin{equation}
\lim_{N\to\infty}\frac{\gamma^{2/(s+2)}}N\rho^{\rm
QM}_{N,a}(\gamma^{1/(s+2)}\x) =\tilde\rho^{\rm TF}_{1,1}(\x)
\end{equation}
in the sense of distributions. Since the functions all have norm
1, we can conclude that there is even weak $L^1$-convergence.

\begin{rem}[The 3D case]
In \cite{LSY2000} the analogues of Theorems \ref{thm11} and \ref{thm12}
were shown for the three-dimensional Bose gas. Using the methods developed
here one can extend these results to analogues of Theorems \ref{thm13} and
\ref{thm14}. In 3D the coupling constant is $g=a$,
so
$\gamma=Na$. Moreover, the relevant mean 3D density is
$\bar\rho_\gamma\sim N (Na)^{-3/(s+3)}$.
\end{rem}

\appendix\section
{Appendix: Scattering length in two dimensions}

Due to the logarithmic behavior of the Green function of the two 
dimensional Laplacian the definition of the scattering length is slightly 
more delicate in two dimensions than in three.  For a nonnegative 
potential $v(\x)$, depending only on $|\x|$ and with finite 
range $R_{0}$, it is naturally defined by the following variational 
principle:

\begin{thm}\label{energy}
 Let $R>R_0$ and consider the functional
\begin{equation}
\mathcal{E}_R[\phi] = \int_{|\x|\leq R}\left\{ |\nabla \phi(\x)|^2 + 
\frac{1}{2}v(\x)|\phi(\x)|^2\right\}{\rm d}^2\x.
\label{E}
\end{equation}
Then, in the subclass of functions such that $\int(|\phi|^2+|\nabla 
\phi|^2)<\infty$ and  $\phi(\x) =1$ for 
$|\x|=R$,
there is a unique function $\phi_0$ that minimizes $\mathcal{E}_R[\phi]$.
This function is nonnegative and rotationally symmetric, and satisfies
the equation
\begin{equation}\label{dist}
-\Delta \phi_0(\x) + \frac{1}{2}v(\x)\phi_0(\x) =0
\end{equation}
for 
$|\x|\leq R$ in the sense of distributions, with  boundary
condition $\phi_0(\x)=1$ for $|\x|=R$.

For $R_0<|\x|<R$
\begin{equation}
\phi_0(\x) =  
\ln (|\x|/a)/\ln(R/a) 
\end{equation}
for a unique number $a$ called the {{\rm scattering length}}.
\end{thm}

For the proof see \cite{LY2000}, where generalizations to other dimensions
and 
potentials with a negative part are also discussed. Note that the factor 
$\frac{1}{2}$ in (\ref{E}) and (\ref{dist}) is due to the reduced mass of the 
two body problem.

If $v$ has infinite range it is easy to extend the definition of the 
scattering length for nonnegative $v$ 
under the assumption that $\int_{|\x|\geq 
R_1}^{\infty} v(\x){\rm d}^2\x <\infty$ for some $R_{1}$.  In fact, one
may 
then simply cut off the potential at some point $R_0 >R_1$ (i.e., set 
$v(\x)=0$ for $|\x|>R_0$) and consider the limit of the scattering 
lengths of the cut off potentials as $R_{0}\to\infty$. See \cite{LY2000}
for 
details.

\end{document}